# Radiation Hardness and Defects Activity in PEA$_2$PbBr$_4$ Single Crystals

Andrea Ciavatti[1*], Vito Foderà[1], Giovanni Armaroli[1], Lorenzo Maserati[1], Beatrice Fraboni[1], Daniela Cavalcoli[1]

[1]Department of Physics and Astronomy – Alma Mater Studiorum University of Bologna, Viale Berti-Pichat 6/2, 40127 Bologna, Italy

*Corresponding Author

**Abstract**

Metal halide perovskites (MHPs) are low-temperature processable hybrid semiconductor materials with exceptional performances that are revolutionizing the field of optoelectronic devices. Despite their great potential, commercial deployment is hindered by MHPs lack of stability and durability, mainly attributed to ions migration and chemical interactions with the device electrodes. To address these issues, 2D layered MHPs have been investigated as possible device interlayers or active material substitutes to reduce ion migration and improve stability. Here we consider the 2D perovskite PEA$_2$PbBr$_4$ that was recently discussed as very promising candidate for X-ray direct detection. While the increased resilience of PEA$_2$PbBr$_4$ detectors have already been reported, the physical mechanisms responsible for such improvement compared to the standard "3D" perovskites are not still fully understood. To unravel the charge transport process in PEA$_2$PbBr$_4$ crystals thought to underly the device better performance, we adapted an investigation technique previously used on highly resistive inorganic semiconductors, called photo induced current transient spectroscopy (PICTS). We demonstrate that PICTS can detect three distinct trap states (T1, T2, and T3) with different activation energies, and that the trap states evolution upon X-ray exposure can explain PEA$_2$PbBr$_4$ superior radiation tolerance and reduced aging effects. Overall, our results provide essential insights into the stability and electrical characteristics of 2D perovskites and their potential application as reliable and direct X-ray detectors.

**Introduction**

The outstanding rise of metal halide perovskites (MHPs), driven by their photosensitive properties, impacts a broad class of opto-electronic semiconducting devices like photovoltaic solar cells, lasers, LEDs, and visible and X-ray photodetectors. MHPs obtained great performance in terms of power conversion efficiency (PCE) in solar cells, and of sensitivity in photodetectors. However, stability and durability issues arise from moving of ions in the perovskite layer under an external electric field (so called "ion migration") and to their accessible chemical reactions with metal electrodes and environment. To mitigate this aspect of MHPs, 2D layered MHPs are being investigated as alternative or additive materials to improve the stability of MHPs-based devices. 2D MHPs have been studied to stabilize solar cells in 2D single layer[1], as 2D/3D bilayer[2,3], or as passivating layer in 2D/3D mixtures[4]. Among them, the Ruddlesden-Popper halide perovskites with phenylethylammonium (PEA$^+$ = C$_6$H$_5$C$_2$H$_4$NH$_3^+$) as ionic cation have proven to be stable in high-vacuum environment and under X-rays[5,6], and to be an excellent active material for UV photodetectors[7]. Thanks to these peculiar properties highly sensitive and ultra-stable direct X-ray detectors based on PEA$_2$PbBr$_4$ perovskite thin films have been realized and proven to effectively work after 80 days in air and after a radiation stress test of about 4 Gy[8]. Even more, 2D perovskites proved their potential as stable X-ray direct detector in several formulation and device geometry.

Furthermore, low-dimensional perovskites have been reported as direct radiation detectors. Quasi-2D perovskite with formulation BA$_2$MA$_2$Pb$_3$I$_{10}$ and BA$_2$MA$_4$Pb$_5$I$_{16}$ (BA$^+$ = C$_4$H$_{12}$N$^+$, butylammonium) are stable for tens of hours under bias and ionizing radiation both in thick[9] or thin[10] films. Another

quasi-2D PEA$_2$MA$_8$Pb$_9$I$_{28}$ polycrystalline thick film showed very high sensitivity (10 860 μC Gy$^{-1}$ cm$^{-2}$) and low detection limit[11]. As ordered system, single crystals of (F-PEA)$_2$PbI$_4$ have very high resistivity >10$^{12}$ Ω cm, high stability, and low noise thanks to suppressed ion migration due to supramolecular electrostatic interaction between electron-deficient F atoms with neighbour benzene rings.[12] The vast majority of studies on 2D perovskites are focused on the device performance and on the effect to the device signal stability (e.g., sensitivity, dark current, etc...) on aging, environmental stability and radiation hardness at low fluxes (few Grays). The few fundamental physical properties investigations target the materials' optical characteristics (absorption and photoluminescence) aiming at photovoltaic applications.[13] .

In this context, limited knowledge is available on the role and behaviour of electrically active defects, in particular regarding their evolution with aging and under ionizing radiation. Shallow and deep in-gap electronic levels are of fundamental importance for the understanding and control of semiconducting properties of materials. In-gap states investigation on inorganic semiconductors in the last 50 years played a major for their technological development[14]. Deep Levels Transient Spectroscopy (DLTS) and Photo-Induced Current Spectroscopy (PICTS) are powerful techniques for the study of these electrically active defects in semiconductors[15]. In recent years DLTS and PICTS have been employed on lead-halide perovskites[16,17], even if the interpretation of the results is still under debate,[18] and the identification of defects is challenging. Most issues affecting the DLTS/PICTS analyses of 3D perovskites arise from a strong ion migration. This leads to a difficult spectral interpretation that hinders a reliable identification and differentiation of electron traps and migrating ions. Thus, thanks to their superior stability and reduced ion migration, 2D perovskites represent an ideal laboratory to investigate electrically active defects, avoiding most of the concerns of 3D perovskites[19]. Moreover, single crystals, as ordered systems, are more prone to the identification of specific defect states, compared to polycrystalline or 2D/3D blends. Finally, metal-Br perovskite, are expected to have a smaller intrinsic charge carrier density (larger bandgap) compared to metal-I ones, and to be less prone to oxidation by environmental $O_2$, thus further improving time and radiation stability.

Here we investigate the role and evolution of electrically active defects in the prototype 2D perovskite $PEA_2PbBr_4$. To achieve this, we do not employ DLTS due its limitations on high resistivity semiconductors[20], and instead we demonstrate the applicability of PICTS to 2D perovskites. Millimetre-size $PEA_2PbBr_4$ single crystals were synthesized, and their quality demonstrated through UV and X-ray photo-response performance as-grown and after 1 month. Through PICTS we obtained stable and reproducible spectra. Three deep levels were clearly identified, and their evolution under 200 Gy of X-ray and over 1 year investigated and discussed.

**Results and Discussions:**

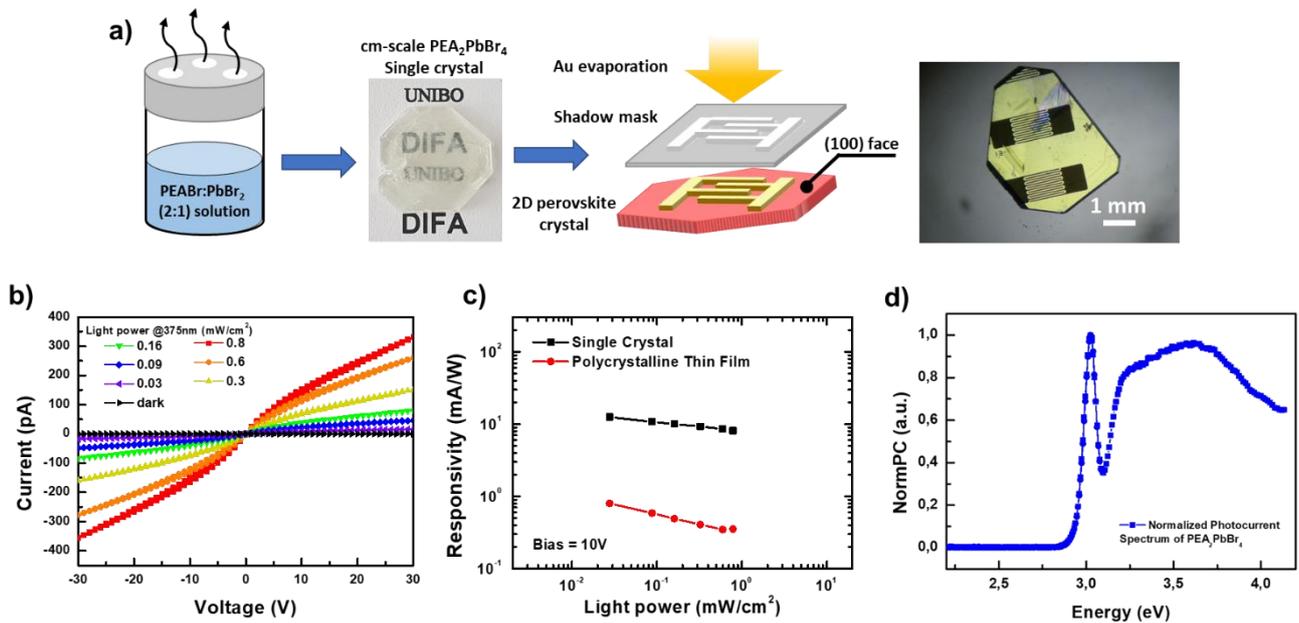

**Figure 1**. Device fabrication and optoelectronics properties. a) Sketch of device fabrication from crystal growth to electrode's evaporation. b) Current-Voltage characteristics of $PEA_2PbBr_4$ single crystals in dark and under 375 nm LED light. c) Responsivity- comparison between free-standing single crystal and spin-coated polycrystalline film. UV-Vis Photocurrent spectrum. d) Normalized UV-vis photocurrent spectrum $PEA_2PbBr_4$ single crystals where large exciton peak and band edge are visible.

The growth of PEA$_2$PbBr$_4$ single crystals follows the slow solvent evaporation technique. Starting from the precursors PEABr and PbBr$_2$ the following reaction proceeds:

$$2\ PEABr + PbBr_2 \rightarrow PEA_2PbBr_4$$

The precursors, mixed in stochiometric fractions, are dissolved in DMF (N,N-Dimethylformamide) solvent to form a 1:3 molar solution (slightly above the supersaturation value of the solute in the solvent). The solution is passed through 0.22 µm filters and put to rest in a beaker covered with Parafilm® where small holes of about one millimetre of diameter were punched. Over days, as the solvent evaporation continues, the solute molarity passes the saturation threshold causing a single crystal to nucleate. After about three weeks, the PEA$_2$PbBr$_4$ crystal is grown to an approximately octagonal shape with sides ranging from a few millimetres to about one centimetre.

Interdigitated metal electrodes are deposited through shadow mask by thermal evaporation of gold or chromium, with channel length L = 30 µm and total width of W = 18.23 mm (Figure 1a). In this architecture, the crystal behaves as photoconductor with co-planar electrodes. When voltage is applied to the electrodes, an electric field is present along the high-conductivity in-plane direction (i.e. the direction of inorganic PbBr$_6$ layers). Current-Voltage characteristics in dark and under the excitation of a 375 nm LED source are reported in Figure 1b. The metal electrodes form an ohmic contact with the crystal, with a very high resistivity of about $10^{12}$ Ω [21], granting an extremely low dark current of few pA at 10V. Despite the large exciton binding energy, the photocurrent response under UV illumination (LED at 375nm) is very efficient, showing an on/off current ration up to $10^4$ at 0.8 mW/cm$^2$ and a signal-to-noise ratio (SNR) equal to 440 already at 30 µW/cm$^2$. The Responsivity reaches 12 mA/W, in accordance with pure PEA$_2$PbBr$_4$ single crystals[7] and 20 times higher than the responsivity of polycrystalline PEA$_2$PbBr$_4$ films grown by spin coating (Figure 1c)[8]. These values confirm the good quality of the crystals used in this study. In the photocurrent spectrum of Figure 1d two high signal regions are identified: the large exitonic peak at 3.02 eV and the continuous band. Appling the Elliott Formula[22,23] at the continuous region in the vicinity of the threshold, the crystals result with an energy gap $E_g$ = 3.08 eV and the excitonic peak about 0.06 eV below the band gap. The temperature-dependent electrical conductivity in the 200 – 340 K range follows the Nernst-Einstein equation, with one single activation energy (Figure SI 1a,b) equal to 0.11 eV, meaning that only one crystalline phase is present in all the considered temperature range. On the contrary, 3D perovskites like $MAPbBr_3$ typically undergo crystal phase transitions and present two (or more) temperature ranges with different activation energies (Figure SI 1c). In particular, high activation energies are extracted at high temperatures, associated to ionic transport.[24,25]
Activation of ionic transport have been observed in $PEA_2PbBr_4$ only at temperature above 340 K[19], much higher that 3D perovskite counterparts, indicating that close room temperature (or below) the ionic transport is strongly reduced.

The response of the devices under X-ray have been tested at different voltages (5 – 50 V) and at different dose rates (2.5 – 25 µGy/s) under an W-target X-ray tube between 40 -150 kVp, see Figure 2a,b. The crystals showed linear response to increasing dose rates for all applied biases, reaching a maximum of sensitivity of 64.8 ± 0.7 µC $Gy^{-1}$ $cm^{-2}$ at 50 V. Figure 2d reports the X-ray photocurrent response spanning over three orders of magnitude of dose rate, from 2.5 µGy/s to 1330 µGy/s. Extending the range to lower dose rates, a minimum detectable dose of 200 nGy/s have been calculated (Figure SI 2). To evaluate the mobility-lifetime product ($\mu\tau$) we acquire the signal output from the detector under the irradiation of an alpha emitting source of $^{226}Ra$. The α-particles have a low-penetration depth in solid state materials, and they stop within few tens of microns from the impinging surface of the perovskite crystal. Due to interdigitate geometry of the electrodes and to the layered structure of the 2D perovskite crystals, the charges are collected only close to the top surface where the metal electrodes were deposited. The use of α-radiation allows to achieve full charge collection and to reach the saturation of photocurrent with he applied bias. The last condition is fundamental for the correct application of the Hecht formula[26] and it would not have been get using penetrating X-ray radiation. Figure 2c reports the experimental data and the corresponding Hecht fit, resulting in a µτ value of $(3.2 \pm 0.3) \times 10^{-6}$ cm$^2$V$^{-1}$.

2D perovskites are generally considered stable materials to environment and others physical stress (like bias or radiation effects). The (PEA)$_2$PbBr$_4$ single crystals presented in this work exhibited very good stability to radiation (Figure 2e) and aging (Figure 2f). Figure 2e reports the electrical resistance in dark conditions and the X-ray sensitivity before and after the exposition to 200 Gy or X-ray at 150 kVp: the values are perfectly overlapped within experimental errors. Figure 2f displays X-ray photocurrent and sensitivity for crystals as-fabricated (t = 0) and after 30 days of storage in ambient conditions. Interestingly, after one year of shelf storage, the µτ were evaluated equal to $6.0 \times 10^{-7}$ cm$^2$V$^{-1}$, retaining 20 % of its initial value after such long time and several characterizations that stressed the device. These last results agree and extend our previous work where PEA$_2$PbBr$_4$ films demonstrated to be stable to aging (80 days) and to continuous X-ray irradiation.

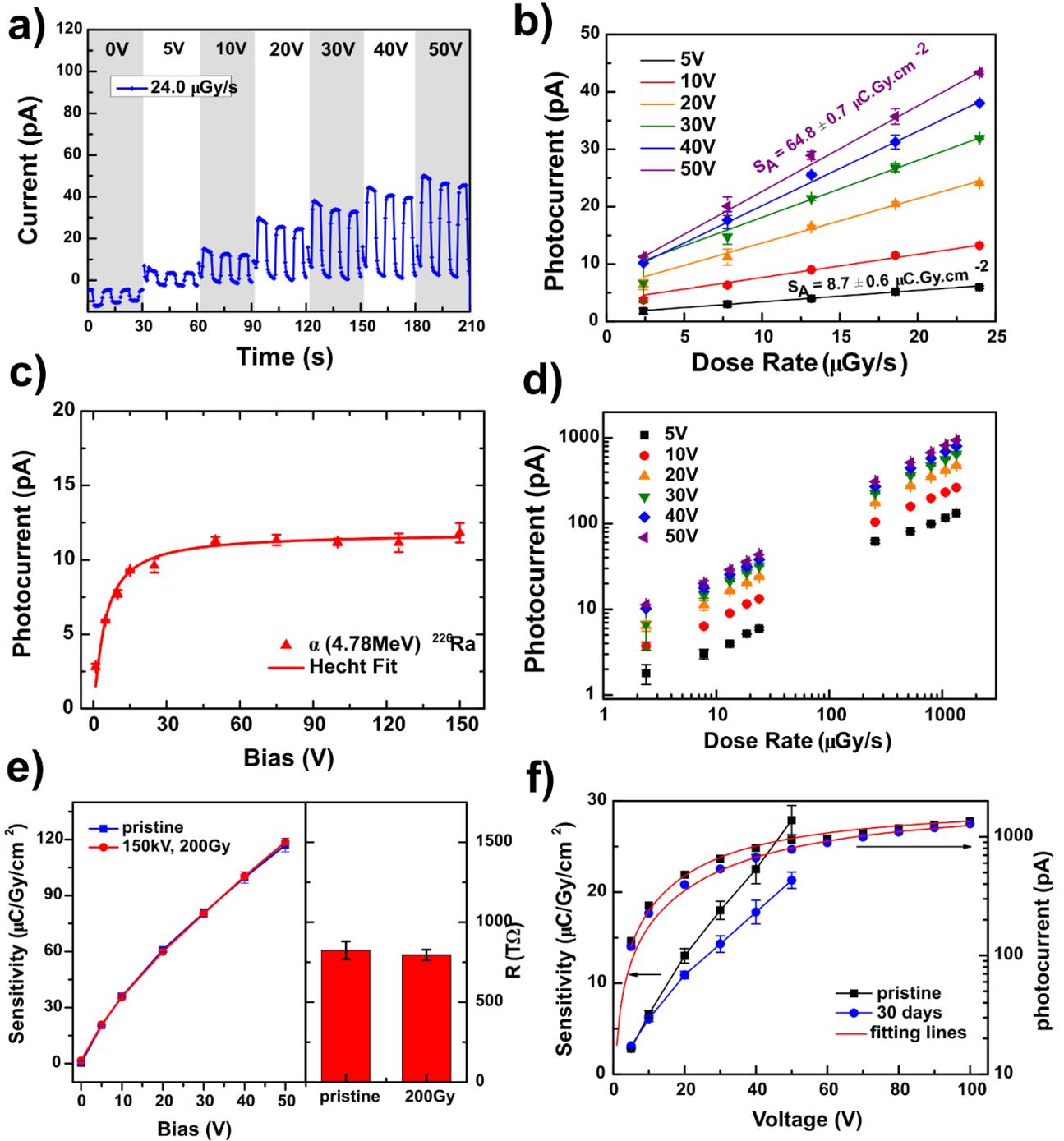

**Figure 2.** X-ray characterization, radiation hardness and aging. a) dynamic X-ray response at fixed dose rate of 24 µGy/s, for bias 5V, 10V, 20V, 30V, 40V, 50V. X-ray source W-target at 40 kVp. b) Linearity of X-ray photocurrent vs. dose rate and corresponding values of X-ray sensitivity. c) Photocurrent vs. bias for alpha-emitting source of $^{226}$Ra (red triangles) and corresponding Hecht fit (solid line). d) X-ray photocurrent vs. dose rate in the range 2.5 – 1330 µGy/s. e) X-ray sensitivity versus applied bias (solid lines, left axis) and detectors resistance (vertical bars, right axis) for pristine crystals (blue) and after 200 Gy of X-ray at 150kV (red). f) X-ray sensitivity (left axis) and photocurrent at 1330 µGy/s (right axis) at 40kVp vs. applied bias for samples as-fabricated (t = 0, black dots) and after 30 days (blue dots).

We have performed Photo-Induced Current Transient Spectroscopy (PICTS) measurements on the PEA$_2$PbBr$_4$ single crystals to investigate the deep energy levels related to charge transport and to follow their evolution over time and under ionizing radiation. PICTS is a spectroscopic technique developed to study electrically active defects in highly resistive photoconductive materials, where the most common Deep Level Transient Spectroscopy (DLTS) is not applicable. In a PICTS experiment, a pulsed optical excitation is used to generate charge carriers inside the material, allowing them to flow and fill intragap trap states. When the optical excitation is switched off, the relaxation transient of the electric current is measured through two ohmic contacts. Optical excitation and transient acquisition are repeated during a temperature scan (typically from 60 K to 450 K), allowing to extract the (thermal) activation energy of each identified deep trap. As the traps thermal emission rate ($e_n$) have a dependence with temperature (T) as follow:

$$e_n(T) = \gamma T^2 \sigma\, e^{-\frac{E_a}{k_B T}} \qquad (1)$$

Where, $\sigma$ is the electron/hole capture cross section, $E_a$ is the level activation energy, and $\gamma$ is a proportionality constant related to charge carriers' effective mass and to conduction/valence band availability. From the full dataset of current transients as function of temperature it is possible to experimentally identify specific levels, access their thermal emission rate $e_n$ (or $e_p$ for holes), and therefore measure the capture cross section, activation energy and, possibly, concentration for each trap present in the material. Figure 3a shows a sketch of the PICTS experimental setup. The perovskite sample is placed inside a cryostat with controlled temperature from liquid nitrogen temperature up to 450 K. A 365 nm LED illuminates the sample through an optical window, with a driving excitation frequency of 13 Hz and optical power of 14 mW/cm$^2$. The sample is biased at 10 V through a low noise current transimpedance amplifier connected to an oscilloscope to display and collect the output signal. In Figure 3b the PEA$_2$PbBr$_4$ output of several current pulses at different temperatures from 79 K to 340 K. The PICTS signal $S$ is obtained from the relaxation transient in Figure 3b applying the rate window concept: we choose two successive instants during the relaxation transient, which start at $t_0$, such as $t_1 > t_0$ and $t_2 > t_1$, and express the PICTS signal as the difference in the value of the currents in these two instants, resulting in:

$$S(T, t_1, t_2) = i(t_1) - i(t_2) \qquad (2)$$

See in Figure 3c. Typical PICTS spectrum resulting for PEA$_2$PbBr$_4$ is reported in Figure 3d in the temperature range from 79 K – 350 K for selected rate windows equal to $(t_2 - t_1)^{-1}$. Each pair of ($t_1$, $t_2$) has a biunivocal relation with emission rate $e_n$ through a non-linear transcendental equation (see SI). If an electrically active trap state is present, thus giving a non-negligible contribution to the photocurrent relaxation transient, a peak in the PICTS spectrum appears, corresponding to a specific emission rate. The emission rate of a trap state has a dependence with temperature T described by equation (1), thus different rate windows are followed by a shift of the $T_{max}$ of each peak. In the PICTS spectra of Figure 3d three peaks can be identified: as no other references for 2D perovskite DLTS/PICTS spectra are known, we labelled them T1, T2 and T3. T1 and T2 are the two dominant peaks, visible in the temperature range between 150 – 180 K and 175 – 200 K, respectively. T3 is the least pronounced peak but clearly emerges at high-rate windows (i.e. high emission rates). Noteworthy, the PICTS spectrum results stable over multiple runs, meaning continuous applied bias and pulsed light in vacuum for several hours and for multiple temperature ramps. The same technique applied to 3D perovskite crystals (like MAPbBr$_3$), where mixed electronic-ionic conduction governs the charge transport, can yield photocurrent transients with much slower response and impedance continuous variation under illumination, due to ionic charges redistribution[19]. The $T_{max}$ and the corresponding $e_n$ are plotted in the Arrhenius plot of Figure 3e where each series of points is associated to a specific trap state. From linear fits in the Arrhenius plot and following equation (1) it is possible to measure the activation energy $E_a$, reported in Table 1. Thermal transient techniques are powerful tools for defects analysis, with high sensitivity even to small concentrations of trap states, however, attention must be paid in the interpretation of peaks in the PICTS spectrum or in the Arrhenius plots and avoid artefacts due to special choice of rate window. To better display the full set of data in one plot, and to verify the reliability of our data, an alternative representation of an Arrhenius plot is presented in Figure 3f. Here a 2D colormap $\ln(e_n)$ *vs* 1000/T represents all the possible combination of ($e_n$, T) for the acquired dataset[27]. In this representation, a trap state is characterized by a maximum in the 2D map whose values are stretched from high $e_n$ at high temperature (top left in the plot) to lower $e_n$ at lower temperature (bottom right in the plot). Every other maximum (symmetric peaks, horizontal/vertical lines, reverse shift) is an artefact to be excluded. The PICTS map confirms the correct interpretation of T1, T2 and T3. To summarize, T1 has an activation energy of 0.32 ± 0.05 eV, T2 of 0.39 ± 0.05 eV, and T3 of 0.52 ± 0.08 eV. The latter is present in only one of the two measured crystals. The level T1 could be compatible with an interstitial defect of bromide ($Br_i$) whose theoretical energy is expected at 0.34 eV from valence band ($PEA_2PbBr_4$ has dominant hole transport)[28]. T2 and T3 cannot be assigned at this stage. We will also see later that T3 is the most sensitive to external stresses.

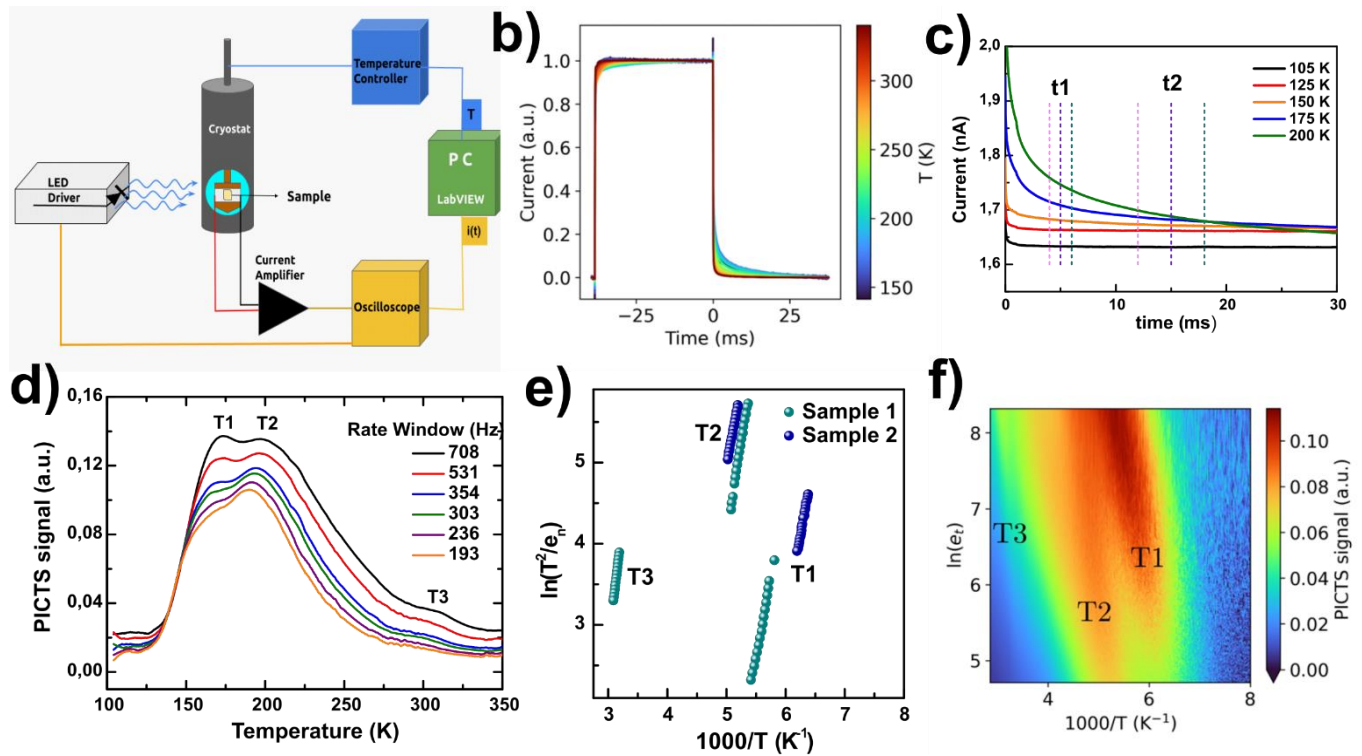

**Figure 3**. PICTS: identification of trap states. a) experimental setup for PICTS measurements. b) Normalized photocurrent pulses at different temperature at 13 Hz for PEA2PbBr4 crystal. c) Example of rate window applied to current transient of plot (b). d) PICTS spectra for selected rate window of 708, 531, 354, 303, 236 and 193 Hz in the range 100 – 350 K. e) Arrhenius plot for peak T1, T2 and T3 identified in PICTS spectra of (d). f) 2D colormap of ln($e_n$) vs. 1000/T as an alternative representation of Arrhenius plot.

| $E_a$ (eV) | T1 | T2 | T3 |
|---|---|---|---|
| **Sample 1** | 0.33 ± 0.05 | 0.40 ± 0.05 | 0.52 ± 0.08 |
| **Sample 2** | 0.31 ± 0.05 | 0.38 ± 0.05 | n/d |

Table 1. Activation energies of traps states from Arrhenius plot.

After having assessed that PICTS can successfully characterize deep levels in 2D perovskites, we used this technique to investigate radiation tolerance and environmental effects on traps states, that can be correlated with the observed macroscopic stability in the device performance in Figure 2e,f.

The ionizing radiation effects on defect levels in Sample 1 have been evaluated using the following protocol. i) A first PICTS measurement has been performed on pristine crystal. ii) a second measurement was repeated after about 2 weeks to assess the repeatability of the sample. iii) Immediately after the second run, the sample was irradiated with a dose of 200 Gy of X-ray at 150 kVp, and then tested again with a third PICTS run. A direct comparison of three PICTS spectra taken for each one of the above steps (identical rate window) and the corresponding Arrhenius plots are shown in Figures 4a and 4b. The complete spectra evolution for various rate windows and corresponding PICTS maps are reported in Figure SI 3. Fa,b show a substantial stability of peaks T1 and T2 even after strong X-ray irradiation, while T3 becomes more prominent after X-ray irradiation. Levels T1 and T2 have completely overlapping Arrhenius plots and their activation energy is very stable both over two weeks of ambient storage and after X-ray irradiation (Table 2). On the other hand, the $E_a$ of level T3 shift of 0.1 eV after a two-week storage, and dramatically changes under ionizing radiation, presenting a very pronounced peak with $E_a$ = 2.2 eV (Figure 4a,b). Such strong modification could suggest that an additional state has been activated. Noteworthy, the absolute photocurrent saturation value under light varies between pristine and irradiated samples, as the quotient PICTS/saturation current is proportional to relative trap concentrations[29]. Figure 4b emphasises that, while the energy levels of traps remain unchanged, the concentration of traps is affected by X-rays with a generalized reduction of relative traps concentration. The photocurrent transient is faster in the irradiated samples (Figure SI 4), that is generally associated with a lower presence of trap states. The reduction is more pronounced for T1 and T2, while T3 emerges from the irradiated PICTS spectrum in Figure 4b.

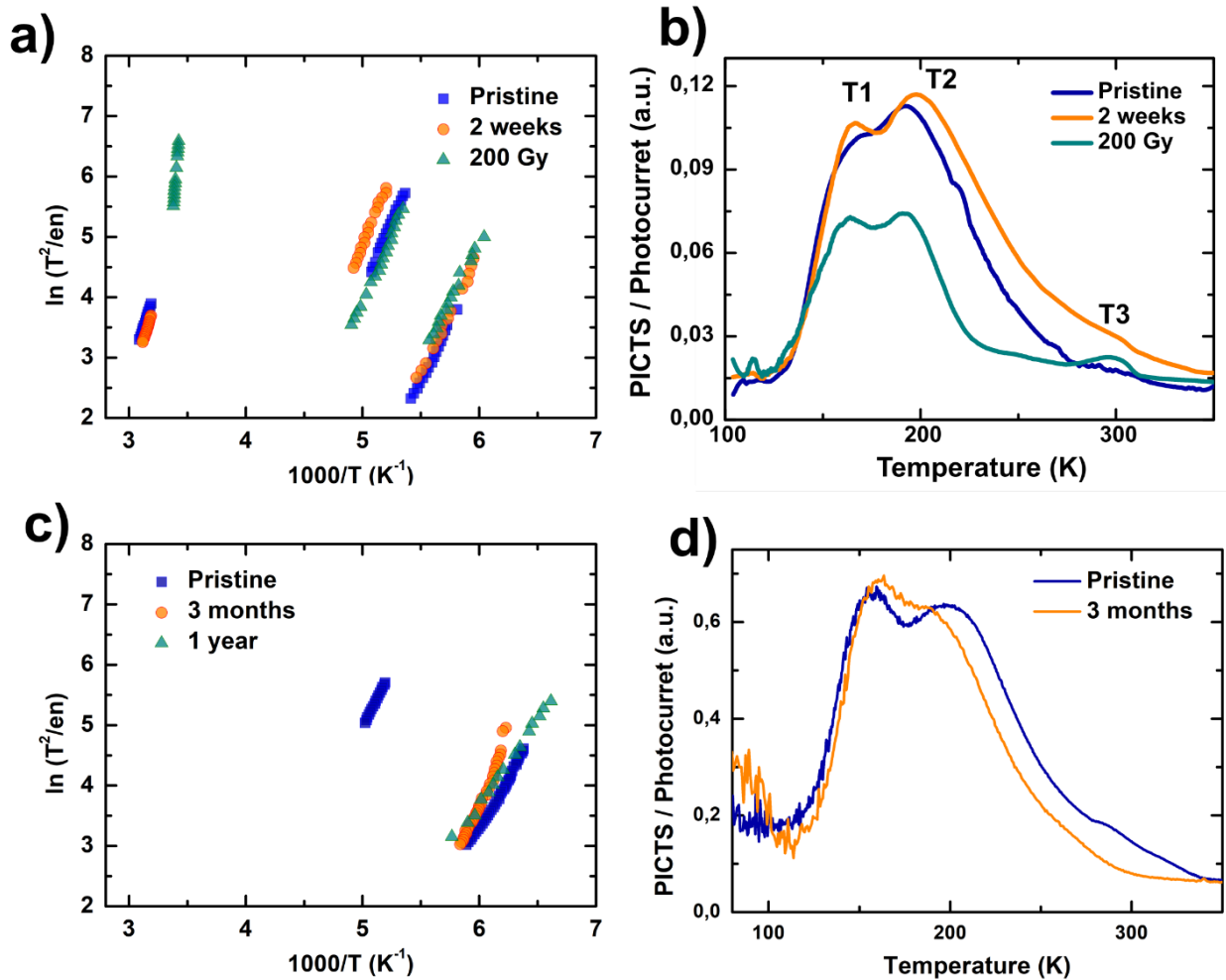

**Figure 4**. Trap states stability: radiation tolerance and aging. a) Arrhenius plot of sample 1 for pristine (blue), after 2 weeks (orange) and after 200 Gy of X-ray (green). b) PICTS signal/photocurrent at 265 Hz for the plot in (a). c) Arrhenius plot of sample 2 for pristine (blue), after 3 months (orange) and after 1 year (green). d) PICTS signal/photocurrent vs. temperature for aging experiment at 256 Hz. This value is proportional to the relative trap concentration.

In *Sample 2*, we followed the evolution of PICTS spectrum during a period of one year to evaluate the aging effects. The samples were stored in standard ambient condition during the whole year. Figure 4c,d report the Arrhenius plots and the relative traps concentration (PICTS/photocurrent spectrum) for pristine crystal, after 3 months and after 1 year. The level T1 results to be the most stable, having the data points in the Arrhenius plot of Figure 4c perfectly superimposed and no difference in trap concentrations. The level T2 slightly reduces its relative concentration after 3 months, while it is stable in the activation energy. Overall, as observed under X-rays, the level T1 and T2 result to be the traps mainly related to material intrinsic properties, that are stable over 3 months. T3 has variable behaviour, it is present in lower concentration with respect to T1 and T2 and, importantly, plays a minor role in the detector performance. Considering the faint and nonconstant presence of T3 in the PICTS spectra, and the difficulties in extracting reliable trap state cross section associate to it, we refrain from drawing conclusive claims on the nature of such electronic state, leaving this task open to further investigation.

| Ea (eV)   | T1            | T2            | T3            |
|-----------|---------------|---------------|---------------|
| **Pristine**  | $0.33 \pm 0.05$ | $0.40 \pm 0.05$ | $0.52 \pm 0.08$ |
| **Two weeks** | $0.35 \pm 0.05$ | $0.40 \pm 0.05$ | $0.62 \pm 0.08$ |
| **X-rays**    | $0.35 \pm 0.05$ | $0.39 \pm 0.05$ | $2.2 \pm 0.3$   |

Table 2: Activation energies of levels T1, T2 and T3 calculated from Figure 4.

**Conclusions**

We grew 2D layered $PEA_2PbBr_4$ perovskite single crystals as platform to study electrically active defect states. First, we showed that they are promising materials for UV-Vis and reliable real-time direct X-ray detectors from 40kVp to 150kVp, thanks to their very low dark current, high quality optoelectronic properties and stability to aging and radiative environment. Then, we used PICTS technique on the 2D perovskite crystals to investigate the defect states. Three levels have been identified: i) T1 at 0.32 eV that we tentatively assigned to Br interstitials (based on ref.[28]); ii) T2 at 0.40 eV still unknown origin; iii) T3 variable in large range 0.52 – 2.2 eV, that could be related to interaction extrinsic factors to be further investigated. Overall, T1 and T2 showed great robustness both to 200 Gy of X-ray radiation at 150kVp and to aging up to 1 year.

**References**

[1] H. Tsai, W. Nie, J.-C. Blancon, C.C. Stoumpos, R. Asadpour, B. Harutyunyan, A.J. Neukirch, R. Verduzco, J.J. Crochet, S. Tretiak, L. Pedesseau, J. Even, M.A. Alam, G. Gupta, J. Lou, P.M. Ajayan, M.J. Bedzyk, M.G. Kanatzidis, and A.D. Mohite, "High-efficiency two-dimensional Ruddlesden–Popper perovskite solar cells," Nature **536**(7616), 312–316 (2016).

[2] T. Campos, P. Dally, S. Gbegnon, A. Blaizot, G. Trippé-Allard, M. Provost, M. Bouttemy, A. Duchatelet, D. Garrot, J. Rousset, and E. Deleporte, "Unraveling the Formation Mechanism of the 2D/3D Perovskite Heterostructure for Perovskite Solar Cells Using Multi-Method Characterization," J. Phys. Chem. C **126**(31), 13527–13538 (2022).

[3] S. Sidhik, Y. Wang, M. De Siena, R. Asadpour, A.J. Torma, T. Terlier, K. Ho, W. Li, A.B. Puthirath, X. Shuai, A. Agrawal, B. Traore, M. Jones, R. Giridharagopal, P.M. Ajayan, J. Strzalka, D.S. Ginger, C. Katan, M.A. Alam, J. Even, M.G. Kanatzidis, and A.D. Mohite, "Deterministic fabrication of 3D/2D perovskite bilayer stacks for durable and efficient solar cells," Science **377**(6613), 1425–1430 (2022).

[4] A.H. Proppe, A. Johnston, S. Teale, A. Mahata, R. Quintero-Bermudez, E.H. Jung, L. Grater, T. Cui, T. Filleter, C.-Y. Kim, S.O. Kelley, F. De Angelis, and E.H. Sargent, "Multication perovskite 2D/3D interfaces form via progressive dimensional reduction," Nat Commun **12**(1), 3472 (2021).

[5] Y.J. Hofstetter, I. García-Benito, F. Paulus, S. Orlandi, G. Grancini, and Y. Vaynzof, "Vacuum-Induced Degradation of 2D Perovskites," Frontiers in Chemistry **8**, (2020).

[6] T.L. Leung, I. Ahmad, A.A. Syed, A.M.C. Ng, J. Popović, and A.B. Djurišić, "Stability of 2D and quasi-2D perovskite materials and devices," Commun Mater **3**(1), 1–10 (2022).

[7] Y. Zhang, Y. Liu, Z. Xu, H. Ye, Q. Li, M. Hu, Z. Yang, and S. (Frank) Liu, "Two-dimensional $(PEA)_2PbBr_4$ perovskite single crystals for a high performance UV-detector," J. Mater. Chem. C **7**(6), 1584–1591 (2019).

[8] F. Lédée, A. Ciavatti, M. Verdi, L. Basiricò, and B. Fraboni, "Ultra-Stable and Robust Response to X-Rays in 2D Layered Perovskite Micro-Crystalline Films Directly Deposited on Flexible Substrate," Advanced Optical Materials **10**(1), 2101145 (2022).

[9] H. Tsai, S. Shrestha, L. Pan, H.-H. Huang, J. Strzalka, D. Williams, L. Wang, Lei.R. Cao, and W. Nie, "Quasi-2D Perovskite Crystalline Layers for Printable Direct Conversion X-Ray Imaging," Advanced Materials **n/a**(n/a), 2106498 (n.d.).

[10] H. Tsai, F. Liu, S. Shrestha, K. Fernando, S. Tretiak, B. Scott, D.T. Vo, J. Strzalka, and W. Nie, "A sensitive and robust thin-film x-ray detector using 2D layered perovskite diodes," Science Advances **6**(15), eaay0815 (2020).

[11] X. He, M. Xia, H. Wu, X. Du, Z. Song, S. Zhao, X. Chen, G. Niu, and J. Tang, "Quasi-2D Perovskite Thick Film for X-Ray Detection with Low Detection Limit," Advanced Functional Materials **n/a**(n/a), 2109458 (n.d.).

[12] H. Li, J. Song, W. Pan, D. Xu, W. Zhu, H. Wei, and B. Yang, "Sensitive and Stable 2D Perovskite Single-Crystal X-ray Detectors Enabled by a Supramolecular Anchor," Advanced Materials **n/a**(n/a), 2003790 (n.d.).

[13] Y. Chen, Y. Sun, J. Peng, J. Tang, K. Zheng, and Z. Liang, "2D Ruddlesden–Popper Perovskites for Optoelectronics," Advanced Materials **30**(2), 1703487 (2018).

[14] D.V. Lang, "Deep-level transient spectroscopy: A new method to characterize traps in semiconductors," Journal of Applied Physics **45**(7), 3023–3032 (1974).

[15] P. Blood, and J.W. Orton, *The Electrical Characterization of Semiconductors: Majority Carriers and Electron States* (Academic Press, 1992).

[16] V. Pecunia, J. Zhao, C. Kim, B.R. Tuttle, J. Mei, F. Li, Y. Peng, T.N. Huq, R.L.Z. Hoye, N.D. Kelly, S.E. Dutton, K. Xia, J.L. MacManus-Driscoll, and H. Sirringhaus, "Assessing the Impact of Defects on Lead-Free Perovskite-Inspired Photovoltaics via Photoinduced Current Transient Spectroscopy," Advanced Energy Materials **11**(22), 2003968 (2021).

[17] J. A. Peters, Z. Liu, M. C. De Siena, M. G. Kanatzidis and B. W. Wessels, "Defect levels in CsPbCl3 single crystals determined by thermally stimulated current spectroscopy." Journal of Applied Physics 132, 035101 (2022).

[18] M.H. Futscher, and C. Deibel, "Defect Spectroscopy in Halide Perovskites Is Dominated by Ionic Rather than Electronic Defects," ACS Energy Lett. **7**(1), 140–144 (2022).

[19] G. Armaroli, L. Maserati, A. Ciavatti, P. Vecchi, A. Piccioni, M. Foschi, V. Van der Meer, C. Cortese, M. Feldman, V. Foderà, T. Lemercier, J. Zaccaro, J. Guillén, E. Gros-Daillon, B. Fraboni, and D. Cavalcoli, "Photo-Induced Current Transient Spectroscopy on Metal Halide Perovskites: Electron Trapping and Ion Drift," Photo-Induced Current Transient Spectroscopy on Metal Halide Perovskites: Electron Trapping and Ion Drift, (2023) https://doi.org/10.1021/acsenergylett.3c01429.

[20] J.C. Balland, J.P. Zielinger, C. Noguet, and M. Tapiero, "Investigation of deep levels in high-resistivity bulk materials by photo-induced current transient spectroscopy. I. Review and analysis of some basic problems," J. Phys. D: Appl. Phys. **19**(1), 57 (1986).

[21] "Ultralow Self-Doping in Two-dimensional Hybrid Perovskite Single Crystals | Nano Letters," (n.d.).

[22] R.J. Elliott, "Intensity of Optical Absorption by Excitons," Phys. Rev. **108**(6), 1384–1389 (1957).

[23] G. Armaroli, L. Ferlauto, F. Lédée, M. Lini, A. Ciavatti, A. Kovtun, F. Borgatti, G. Calabrese, S. Milita, B. Fraboni, and D. Cavalcoli, "X-Ray-Induced Modification of the Photophysical Properties of MAPbBr3 Single Crystals," ACS Appl. Mater. Interfaces, (2021).

[24] Y. Lin, Y. Bai, Y. Fang, Q. Wang, Y. Deng, and J. Huang, "Suppressed Ion Migration in Low-Dimensional Perovskites," ACS Energy Lett. **2**(7), 1571–1572 (2017).

[25] X. Xiao, J. Dai, Y. Fang, J. Zhao, X. Zheng, S. Tang, P.N. Rudd, X.C. Zeng, and J. Huang, "Suppressed Ion Migration along the In-Plane Direction in Layered Perovskites," ACS Energy Lett. **3**(3), 684–688 (2018).

[26] A. Ciavatti, P.J. Sellin, L. Basiricò, A. Fraleoni-Morgera, and B. Fraboni, "Charged-particle spectroscopy in organic semiconducting single crystals," Applied Physics Letters **108**(15), 153301 (2016).

[27] J.V. Li, "Deep level transient spectroscopy characterization without the Arrhenius plot," Review of Scientific Instruments **92**(2), 023902 (2021).

[28] J. Yin, R. Naphade, L. Gutiérrez Arzaluz, J.-L. Brédas, O.M. Bakr, and O.F. Mohammed, "Modulation of Broadband Emissions in Two-Dimensional ⟨100⟩-Oriented Ruddlesden–Popper Hybrid Perovskites," ACS Energy Lett. **5**(7), 2149–2155 (2020).

[29] A. Cavallini, B. Fraboni, W. Dusi, M. Zanarini, and P. Siffert, "Deep levels and compensation in γ-irradiated CdZnTe," Appl. Phys. Lett. **77**(20), 3212–3214 (2000).

# SUPPLEMENTARY MATERIALS

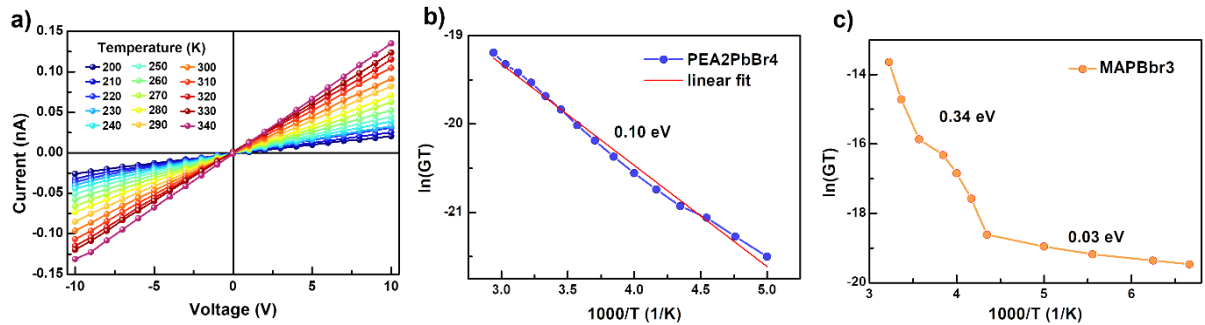

**Figure SI 1**: Temperature dependence of electrical conductivity. a) Current-Voltage characteristics in dark and under vacuum in the temperature range spanning from 200 K to 340 K. 10K step between each plot. b,c) Logarithm of electrical conductivity vs. 1/T showing the linear behaviour of Nernst-Einstein equation for 2D perovskite PEA2PbBr4 (b) and 3D perovskite MAPbBr3 (c).

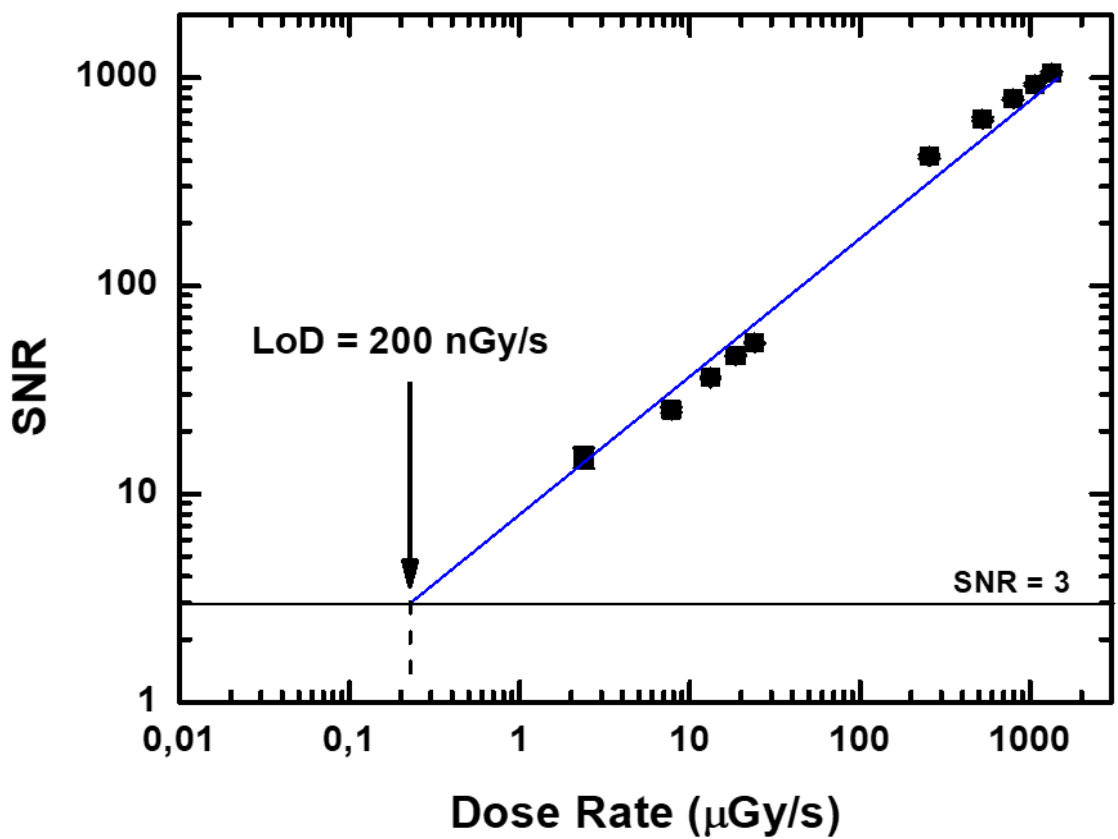

**Figure SI 2**. Signal-to-noise Ratio (SNR) at low dose rates to calculate the Limit of Detction (LoD) under X-ray irradiation. Bias of 10V.

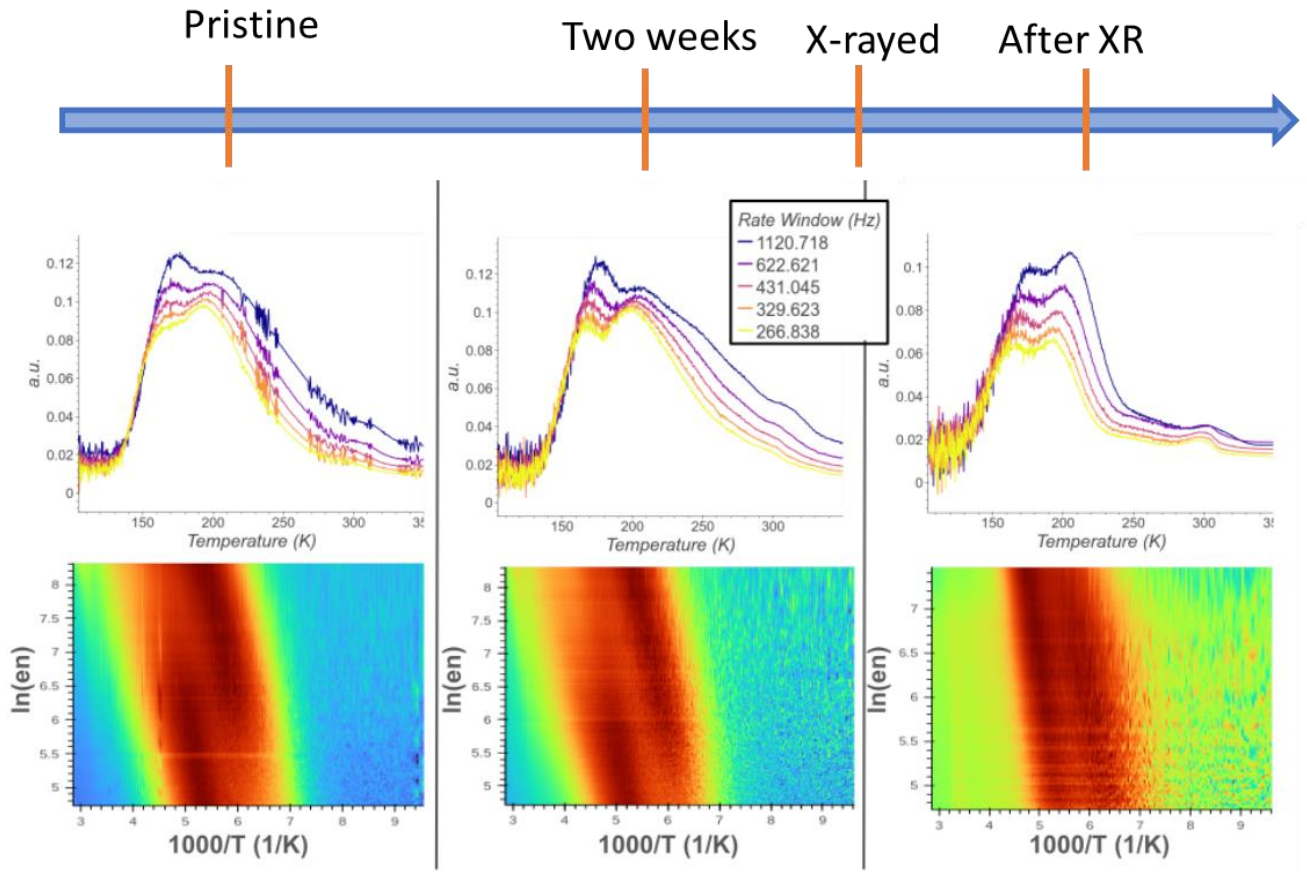

**Figure SI 3**. PICTS under X-ray.

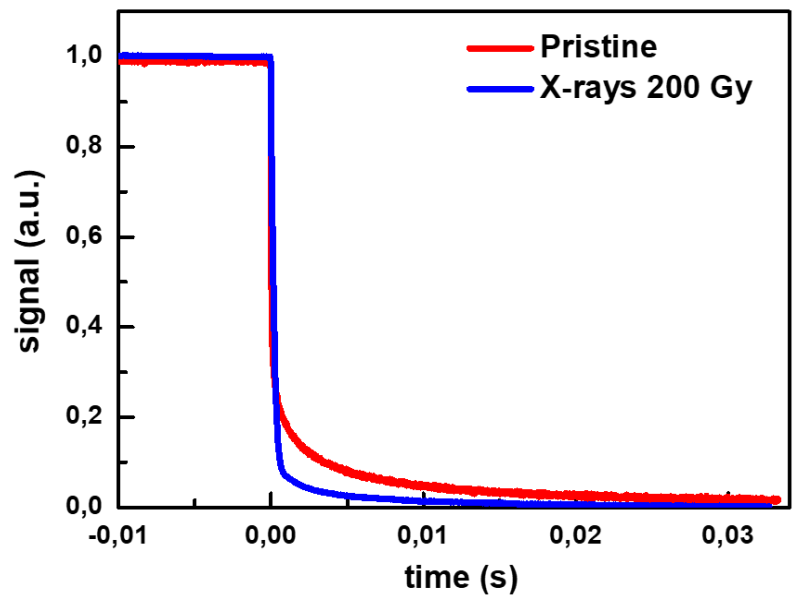

**Figure SI 4**. Collection of current transients as the temperature changes, compared before X-ray exposure and after X-ray exposure.